\documentclass[a4paper,fleqn,usenatbib]{mnras}

\usepackage{newtxtext,newtxmath}

\usepackage[T1]{fontenc}
\usepackage{ae,aecompl}

\usepackage{graphicx}   
\usepackage{amsmath}    
\usepackage{amssymb}    

\title[The SDSS extended PSFs]{The Sloan Digital Sky Survey extended point
spread functions}

\author[Infante-Sainz et al.]{
Ra\'ul Infante-Sainz,$^{1,2}$\thanks{E-mail: rinfante@iac.es;
infantesainz@gmail.com}
Ignacio Trujillo$^{1,2}$
and Javier Rom\'an$^{1,2,3}$
\\
$^{1}$Instituto de Astrof\'isica de Canarias, c/ V\'ia L\'actea s/n,
E-38205, La Laguna, Tenerife, Spain\\
$^{2}$Departamento de Astrof\'isica, Universidad de La Laguna, E-38205 La
Laguna, Tenerife, Spain\\
$^{3}$Instituto de Astrof\'isica de Andaluc\'ia (CSIC), Glorieta de la
Astronom\'ia, E-18008 Granada, Spain\\
}

\date{Accepted 2019 October 30. Received 2019 October 29; in original form
2019 September 10.}

\pubyear{2019}

\newcommand{\projectversion}{v0.5-0-g62d83df}

\newcommand{\nstars}{1050}
\newcommand{\svalueforstar}{0.13}
\newcommand{\svalueforgalaxy}{0.88}
\newcommand{\svaluefordip}{0.07}
\newcommand{\meanslopevalue}{-2.49}
\newcommand{\ufluxexcess}{5.6}
\newcommand{\gfluxexcess}{1.6}
\newcommand{\rfluxexcess}{1.7}
\newcommand{\zfluxexcess}{4.1}

\begin{document}
\label{firstpage}
\pagerange{\pageref{firstpage}--\pageref{lastpage}}
\maketitle

\begin{abstract}
A robust and extended characterization of the point spread function (PSF) is crucial to extract the photometric information produced by deep imaging surveys.
Here, we present the extended PSFs of the Sloan Digital Sky Survey (SDSS), one of the most productive astronomical surveys of all time.
By stacking $\sim$1000 images of individual stars with different brightness, we obtain the bidimensional SDSS PSFs extending over 8 arcmin in radius for all the SDSS filters(\textsl{u}, \textsl{g}, \textsl{r}, \textsl{i}, \textsl{z}).
This new characterization of the SDSS PSFs is near a factor of 10 larger in extension than previous PSFs characterizations of the same survey.
We found asymmetries in the shape of the PSFs caused by the drift scanning observing mode.
The flux of the PSFs is larger along the drift scanning direction.
Finally, we illustrate with an example how the PSF models can be used to remove the scattered light field produced by the brightest stars in the central region of the Coma cluster field.
This particular example shows the huge importance of PSFs in the study of the low-surface brightness Universe, especially with the upcoming of ultradeep surveys, such as the Large Synoptic Survey Telescope (LSST).
Following a reproducible science philosophy, we make all the PSF models and the scripts used to do the analysis of this paper publicly available (snapshot \projectversion).
\end{abstract}


\section{Introduction}
The point spread function (PSF) describes the response of an imaging system to the light produced by a point source.
Real PSFs have complex structures as their shapes depend on the optical path that light takes as it travels through the atmosphere and multiple optical elements, mirrors, lenses, detectors, etc.
For the vast majority of astronomical works, only a tiny portion of the PSF \citep[i.e. normally a few inner arcseconds; see e.g.][]{trujillo_2001a, trujillo_2001b} is characterized.
In practice, however, the light of both point and extended sources are spread over the entire detector due to the effect of the PSF at large radii.
Therefore, it is necessary to have a good understanding of its structure along the entire detector (typically extending over arcminutes or more).

Extended PSFs have become a vital tool to obtain precise photometric information in modern astronomical surveys.
For instance, \cite{slater_2009} modelled the extended PSF and the internal reflections produced by the stars of the Burrell Schmidt telescope and showed that virtually all the pixels of the image are dominated by the scattered light by both stars and galaxies at 29.5 mag/arcsec$^2$ (\textsl{V}-band).
\cite{trujillo_fliri_2016} also characterized and used the extended PSF of the 10.4 m Gran Telescopio Canarias (GTC) telescope to model and remove the scattered light in ultradeep observations of the UGC 00180 galaxy.
Even more troublesome for low-surface brightness studies is the finding \citep[see e.g.][]{trujillo_bakos_2013, sandin_2014,sandin_2015} that the outer regions of astronomical objects are severely affected by their own scattered light produced by the convolution with the PSF.
In order to correct this effect, \cite{karabal_2017} generated the PSF and models of the internal reflections from images of the Canada-France-Hawaii Telescope (CFHT) to de-convolve a sample of three galaxies and correct them from instrumental scattered light.
More recently, \cite{roman_2019} characterized the PSFs of the Stripe 82 survey and used them to model and correct the scattered light field produced by stars to study the optical properties of the Galactic cirri.
All the above works have shown that having an extended PSF is crucial when accurate photometric and structure properties of astronomical objects at low-surface brightness levels are required.

One of the most commonly used surveys for measuring photometric properties of astronomical objects is the Sloan Sky Digital Survey \citep[SDSS;][]{york_2000_sdss_overview}, covering 14 555 deg$^2$ on the sky (just over 35 per cent of the full sky) in five photometric bands (\textsl{u}, \textsl{g}, \textsl{r}, \textsl{i}, and \textsl{z}).
Although SDSS is a relatively shallow survey compared to current \citep[see e.g. IAC Stripe82 Legacy Survey;][]{fliri_trujillo_2016,roman_trujillo_2018}and future wide area surveys \citep[see e.g. Large Synoptic Survey Telescope (LSST), Euclid, Messier;][]{lsst_sciencebook,euclid_2011,messier_2017}, its massive use motivate us to explore their extended PSFs as a useful excercise that can be repeated in future deeper surveys.

Previous attempts to characterize the extended SDSS PSFs were conducted by \citet{zibetti_2004, dejong_2008, bergvall_2010}, and \cite{tal_vandokkum_2011}.
All of them are extensively analysed and discussed in \cite{sandin_2014}.
In general, the goal of these studies was to analyze the outer part of SDSS galaxies in a statistical way to check the role of the scattered light in their low-surface brightness regions.
\cite{dejong_2008} obtained extended \textsl{g}, \textsl{r}, and \textsl{i} SDSS PSFs.
These PSFs have an extension of $\sim$1 arcmin in radius.
Those PSFs were created by the stacking of averaged radial profiles of stars and then assuming (following the shape of the intermediate region) an $\text{r}^{-2.6}$ power-law function to extend their outer parts to a larger distance.
Similarly, \cite{tal_vandokkum_2011} stacked all the bright stars ($8.0<m_r<8.2$) from images of SDSS DR7 \citep{sdss_dr7} in each of the five optical bands to characterize the outer parts of the PSFs, while the inner parts were obtained from non-saturated SDSS PSF models \citep{lupton_2001}.
These authors also reached an extension of $\sim$1 arcmin in radius.
Despite the fact that 1 arcmin is a relatively large extension, a more extended PSF is needed when dealing with nearby galaxies.
In fact, as shown by \cite{sandin_2014}, PSFs of at least 1.5 times the size of objects are needed to make a proper analysis of the effect of the PSF on the light distribution of astronomical sources. In other words, the study of a large number of interesting objects whose size exceeds the 1 arcmin scale requires significantly larger PSFs.
Motivated by this requirement and its utmost importance for studies involving deep imaging analysis, we have characterized the SDSS PSFs down to $\sim$8 arcmin (i.e. approximately an order of magnitude larger in extension than previous works).
These PSFs are made publicly available to the astronomical community (see Sect.~\ref{sec:results}).

This paper is structured as follows.
In Sect.~\ref{sec:psf_building}, we describe all the steps followed in order to build the SDSS extended PSFs.
Section~\ref{sec:results} is dedicated to the characteristics of the PSFs and how they can be used to model and correct for the scattered light of stars in the Coma Cluster SDSS image.
In Sect.~\ref{sec:conclusions}, we provide our discussion and conclusions.

\section{Constructing the extended SDSS PSFs}
\label{sec:psf_building}
The aim of this work is to build, for the SDSS filter set, PSFs as extended as possible.
To achieve this goal, it is necessary to use stars within a wide brightness range.
Very bright ($<$7 mag) stars (although saturated and with bleeding patterns in their central part) are useful for creating the outer part of the PSFs.
The stacking of bright stars ($\sim$9 mag) is done to construct the intermediate region of the PSF, whereas faint ($>$14 mag; non-saturated) stars are used for assembling the central core (inner part).

The stars considered to construct the PSFs are selected from the US Naval Observatory (USNO-B1) Catalog \citep{monet_2003_usno}.
We only take stars having low proper motions (pmRA and pmDEC < 600 mas/yr) and errors in their central spatial coordinates equal to zero (e\_RAJ2000 and e\_DEJ000 = 0).
Each star is selected to construct a different region of the PSF according to its magnitude.
Once the stars from USNO-B1 Catalog are chosen, SDSS images of these stars in each filter are downloaded from the SDSS archive.
We explored the possibility of using Gaia \citep{gaia2_2018} for our selection of the brightest stars, but we found that this catalogue is incomplete for objects with magnitude brighter than 7 mag, and no stars with magnitude below 1.7 mag are present.

The SDSS camera has an array of 30 CCDs (located in 5 rows and 6 columns; each row of the camera corresponds to a filter).
Each CCD has a size of 10 arcmin along the drift scanning direction and 13 arcmin perpendicular to that direction \citep[see][for a detailed description of the SDSS camera]{gunn_1998_sdss_camera}.
We use single SDSS images to create the PSFs.
We avoided using mosaic images generated from the combination of multiple images, because each CCD can have a different sky background value.
None the less, since each star is observed at a different CCD position, it is possible to obtain a PSF larger than the size of single CCDs.
This is because each star will contribute to generate a different region of the extended PSF depending on its location over the CCD image.
Naturally, to have enough coverage of the outermost region of the PSF, it is necessary to have a large number of stars.

For \textsl{g}, \textsl{r}, and \textsl{i} filters, in order to reach a signal-to-noise ratio (S/N) above 3 along the entire radial profiles of the extended SDSS PSFs, we need to combine about 1000 individual star images.
The images in \textsl{u} and \textsl{z} bands are noisier and shallower.
Therefore, the same number of stars only allow us to obtain extended PSF profiles with S/N$\sim$1 at a radius of 8 arcmin for these two filters.
As we are using the brightest stars possible from the SDSS footprint, the addition of fainter stars does not help to get better S/N in the final stacked PSFs.
For the intermediate and inner part of the PSFs we also combine 1000 stars.
In practical terms, we find that in order to have a final number of 1000 useful stars we typically need to explore \nstars{} stars, as some of them are rejected due to quality cuts which will be discussed in next sections.

In order to make our results reproducible, in the next subsections we describe in detail the steps followed to build the PSFs.
The construction of the outer part of the PSFs is described in Sect.~\ref{sec:method_psf_outer}.
The intermediate and inner parts will be briefly described in Sections ~\ref{sec:method_psf_intermediate} and ~\ref{sec:method_psf_core}.
The methodology used for the junction between the inner, intermediate, and outer parts of the PSFs is discussed in Sect.~\ref{sec:method_psf_junction}.

\subsection{Constructing the outer part of the PSFs}
\label{sec:method_psf_outer}
We use the brightest stars from the SDSS footprint in each band to build the outer part of the PSFs.
These stars are not necessarily the same for all the SDSS filters.
In fact, we have selected stars from the USNO-B1 Catalog with brightness < 7.0 mag as followed: \textsl{B1mag} < 7.0 mag for the \textsl{u} SDSS PSF, \textsl{R1mag} < 7.0 mag for the \textsl{g} and \textsl{r} SDSS PSFs, and \textsl{Imag} < 7.0 mag for the \textsl{i} and \textsl{z} SDSS PSFs.
The magnitudes of the stars in \textsl{B1mag} (optical \textsl{B} band between 400 and 500 nm), \textsl{R1mag} (optical \textsl{R} band between 600 and 750 nm), and \textsl{Imag} (optical \textsl{I} band between 750 and 1000 nm) are given in the USNO-B1 Catalog.
This catalogue presents magnitudes in the photographic system (O, E, J, F, N)\footnote{O, E, J, F, N photographic system magnitudes correspond to \textsl{B1mag}, \textsl{B2mag}, \textsl{R1mag}, \textsl{R2mag}, and \textsl{Imag} in the \href{https://vizier.u-strasbg.fr/viz-bin/VizieR?-source=I/284}{USNO-B1 VizieR catalogue.}}, and not in standard systems \citep[see][for a detailed description on how USNO-B1 Catalog was obtained]{monet_2003_usno}.

The construction of the PSF's outer part is as follows.
First, we recentre each individual star into a newer and larger image (im1) with an odd number of pixels in order to have the centre of the star in the central pixel of im1.
To do that, we run \texttt{SWarp} \citep{bertin_2010_swarp} using as the central position of the stars the coordinates provided by the USNO-B1 Catalog.
Aiming to create a model for the outer part of the PSF as a result of the stacking of many individual stars, it is critical to detect and mask all the external sources.
On doing this, we avoid the contribution of this undesired flux to the PSF model.
To implement such a step, we run \texttt{NoiseChisel}\footnote{\texttt{NoiseChisel}, \texttt{Segment}, and other programs like \texttt{Arithmetic} are extensively used in this work. They belong to \href{https://www.gnu.org/software/gnuastro/}{GNU Astronomy Utilities} software.} and \texttt{Segment} \citep[]{akhlaghi_2015_noisechisel, akhlaghi_2019_carving} to create a segmentation map of each re-centred image (im1).
We used the following \texttt{NoiseChisel} and \texttt{Segment} parameters\footnote{The meaning of the parameters used can be found in the \href{https://www.gnu.org/software/gnuastro/manual/}{documentation book} of this software.} to optimize the source detection and segmentation: \texttt{tilesize=100,100}; \texttt{interpnumngb=1}; \texttt{minnumfalse=1}; \texttt{qthresh=0.5}.
The segmentation map is used to mask all flux contributions that does not belong to the central bright star.

To stack all the individual stars, the flux of each  image is normalized to the flux within a ring of 1 pixel width (SDSS pixel scale is 0.396 arcsec pixel$^-1$) at a radial distance of 60 arcsec from the centre of the star.
This radial distance is selected because it is close enough to the centre, therefore a good S/N is obtained, but is also distant enough to avoid the region where most of the internal reflections produced by the optics of the telescope are found.
Each image is divided by the mean flux (n1) of this ring.

\begin{figure}
	\includegraphics[width=\columnwidth]{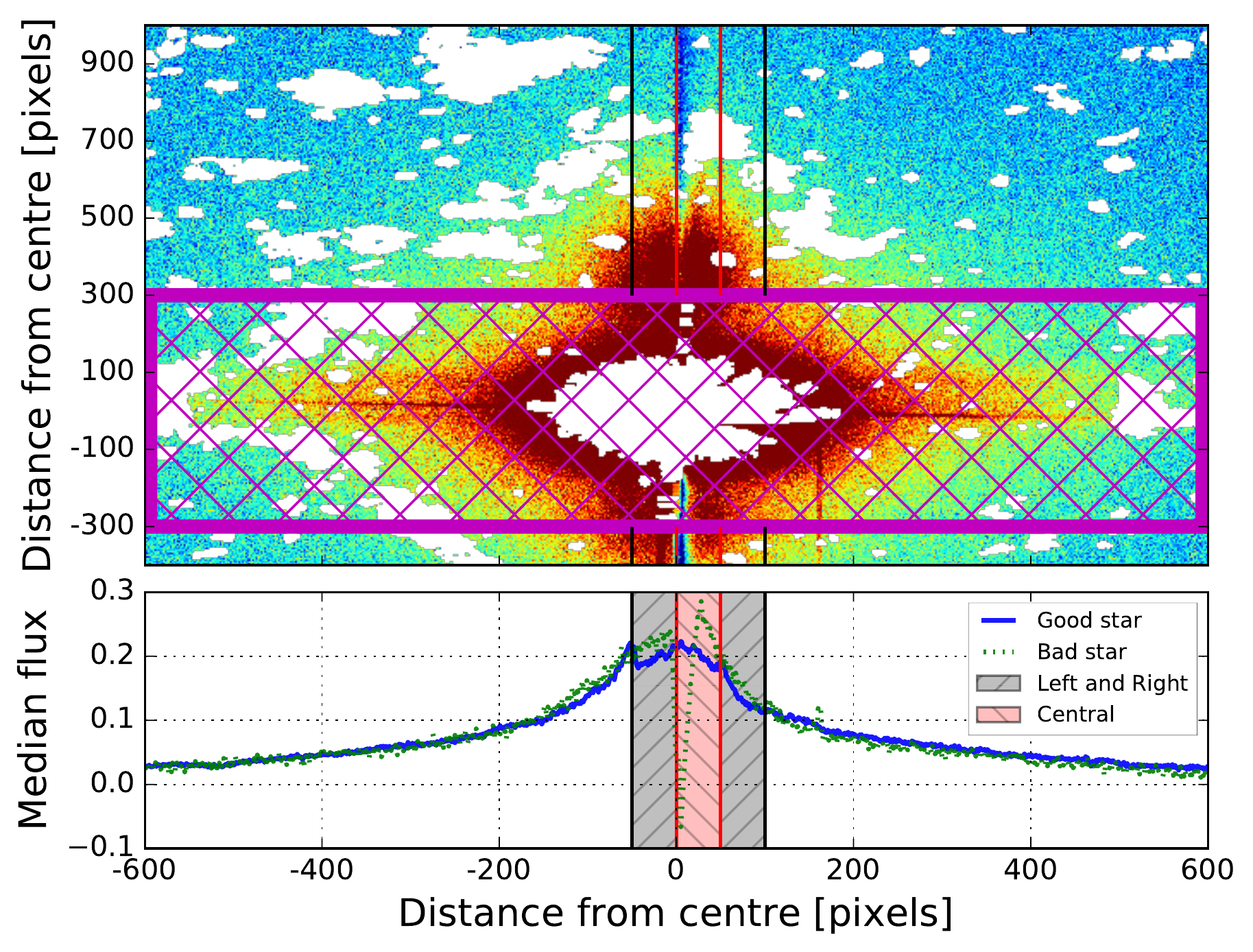}
    \caption{An illustration of how the stars affected  by the depletion in the saturated pixels are rejected. Upper panel: SDSS \textsl{i} band image of a bright star (R.A. (J2000) = 22:55:10.96; Dec (J2000) = -04:59:16.37).
             The drift scanning direction of the survey goes from top to bottom.
             White areas correspond to masked pixels using \texttt{NoiseChisel} and \texttt{Segment} object masks.
             Inaccurately corrected saturated pixels by the SDSS pipeline can be seen above and below the star in the central region as a decrease in the flux of the image.
             The cross lined magenta rectangle encloses pixels that are masked in order to obtain the 1-dimensional profile of the star.
             Bottom panel: collapsed 1-dimensional profiles of two stars perpendicular to the drift scanning direction using the median value.
             With solid blue line, we plot the profile of one star (R.A.(J2000) = 14:19:45.23; Dec (J2000) = 16:18:25.02) unaffected by the problem of the depletion of saturated pixels ($\text{p}=0.84$).
             With dotted green line is plotted the profile of the star of the upper panel with the problem of the depletion ($\text{p}=-14.36$).
             This can be clearly seen as a sharp drop of the profile in the central region. Parameter $\text{p}$ is defined in Eq.~\ref{eq:reject_stars} (see the text for more details).}
    \label{fig:psf_i_reject_satstars}
\end{figure}

Bright stars are severely affected by bleeding and saturated pixels in SDSS images.
The SDSS team has allocated a significant effort to correct this issue.
None the less, the correction of the bleeding and saturated pixels done by SDSS pipeline \citep{stoughton_2002_sdss} still shows (for some of the brightest stars) a depletion along the saturated pixel columns which are not part of the PSF.
This effect becomes visible approximately at a radial distance of 300 pixels from the centre along the drift scanning direction.
To avoid being affected by this issue at creating the PSF models, we reject those stars which have this depletion in the SDSS filters \textsl{g}, \textsl{r}, and \textsl{i}.
In \textsl{u} and \textsl{z} SDSS bands, the correction done by the SDSS pipeline of the saturated pixels is good enough and does not affect the creation of the PSFs.
The depletion is especially relevant in the \textsl{i} band, and because of that, we illustrate this issue in Fig.~\ref{fig:psf_i_reject_satstars}.

The rejection of stars affected by the depletion is done by analysing the pixels of the saturated region and comparing them with the surrounding pixels close to this region.
To do that, first of all, we mask a central band of 600 pixels width perpendicular to the drift scanning direction (cross lined region in the upper panel of Fig.~\ref{fig:psf_i_reject_satstars}).
Then we collapse vertically the image into a 1-dimensional vector using the median values of the non masked pixels.
This 1-dimensional vector is equivalent to an average profile of the outer part of the star along the perpendicular direction to the drift scanning direction (bottom panel of Fig.~\ref{fig:psf_i_reject_satstars}).
The central part of the profile is divided into three regions. Depending on the SDSS filter, these regions are placed at different positions motivated by the location of the depletion.
For \textsl{g} and \textsl{r} SDSS bands the depletion is located in the middle of the image.
Consequently, we put the central region in the middle of the image with a width of 10 pixels.
For the \textsl{i} SDSS band, the depletion is not exactly at the central position, therefore the central region is moved 5 pixels from the centre to the right and has a width of 50 pixels.
The other two regions are located at the left and right sides of the central region, and have 50 pixels width for all bands.
Using these regions, we define the parameter $\text{p}$ as the ratio between the minimum value of the star profile in the left and right regions, and the minimum value of the central region.
That is,

\begin{equation}
    \text{p} = \frac{\min(\text{left},\,\text{right})}{\min(\text{centre})}.
	\label{eq:reject_stars}
\end{equation}

With this definition, if the correction of the saturated pixels is reasonable, $\text{p}$ is expected to be between 0 and 1.
Note that $\text{p}<0$ implies that the profile of the star in the central region has a minimum value with a different sign than the profile values on the regions in each side.
Alternatively, $\text{p}>1$ means that the profile has a strong depletion in the central region compared to the profile in the other two explored regions.
Consequently, we reject all stars with $\text{p}<0$ or $\text{p}>1$ until the accumulated number of acceptable downloaded stars reaches \nstars{} in each band.
For \textsl{r} and \textsl{g} filters, 1 out of 5 stars are rejected because of a strong depletion problem, while in the \textsl{i} band, this issue affects 1 out of 3 stars.
Once we have done all the previous steps, we stack all masked and normalized star images using 3$\sigma$ clipping median\footnote{In general, the median is a much more robust operator than the mean because it is less affected by outlier values. We compared stacked images using mean and median operator with 3$\sigma$ clipping, and the results are quite similar with sigma clipping median operator stacked images having slightly less noise.}.

After the stacking of the brightest SDSS stars, we obtain an image which is a first version of the outer part of the PSF (psf1).
However, due to the brightness of the stars we have used, many contaminant sources (particularly those radially close to the central region of the stars) were not properly deblended during the first run using \texttt{NoiseChisel} and \texttt{Segment}.
For this reason, to improve our masks, we do the following: we subtract the psf1 from each original image (im1) using the normalization value estimated before (n1), leaving a residual image (im2) in which we run \texttt{NoiseChisel} and \texttt{Segment} again to obtain better object masks.
That is,

\begin{equation}
    \text{im2} = \text{im1} - (\text{n1} \times \text{psf1}).
	\label{eq:iteration1}
\end{equation}

\texttt{NoiseChisel} and \texttt{Segment} do the detection of pixels using boxes with a size as the one provided by the \texttt{tilesize} parameter, used to estimate the sky background and the S/N.
For this reason, sometimes the detected pixels have the shape of these tilesizes.
In order to avoid this effect, we choose random tilesizes.
The parameters of \texttt{NoiseChisel} and \texttt{Segment} in the second iteration are \texttt{qthresh=0.5}, \texttt{minnumfalse=1}, \texttt{interpnumngb=1}, and \texttt{tilesize=irandom,irandom}, where \texttt{irandom} is a random integer between 70 and 100.
Thanks to this second iteration, it is possible to mask the internal reflections and the diffraction spikes of stars in a much better way.
With the new masks, the normalization of each image is improved and the stacking of all im2 images (with the improved masks and normalization) results in a much better PSF (psf2) than the initial one (psf1).

By following the previous steps, the outer part of the PSFs have been built stacking about \nstars{} stars in each filter.
However, not all the stars are equally bright.
In fact, this difference in brightness plays a major role in the outermost part of the PSF if we want to create reliable PSFs.
To guarantee that our PSFs are accurate enough along the entire radial range of exploration, we have modified our stacking procedure such at all radial distances the brightness of the profiles of the stars that are contributing to the stacking have brightness above the surface brightness limit of the images.
This is done as follows.
We extract the circularly averaged surface brightness radial profile of each star to estimate at which radial distance ($\text{R}_{mlim}$) the surface brightness profile reaches $\mu$=26.5 mag/arcsec$^2$.
This value is chosen as is the typical surface brightness limit of the SDSS survey \citep[i.e. 3$\sigma$ in 10$\times$10 arcsec$^2$ boxes;][]{pohlen_trujillo_2006, trujillo_fliri_2016}.
Beyond $\text{R}_{mlim}$ the signal of the star is so poor that if we were using these regions, the PSF model would have been affected by the inclusion of noise.
Consequently, we mask all the pixels beyond $\text{R}_{mlim}$ for each star that is used for the final stacking.

\begin{figure}
	\includegraphics[width=\columnwidth]{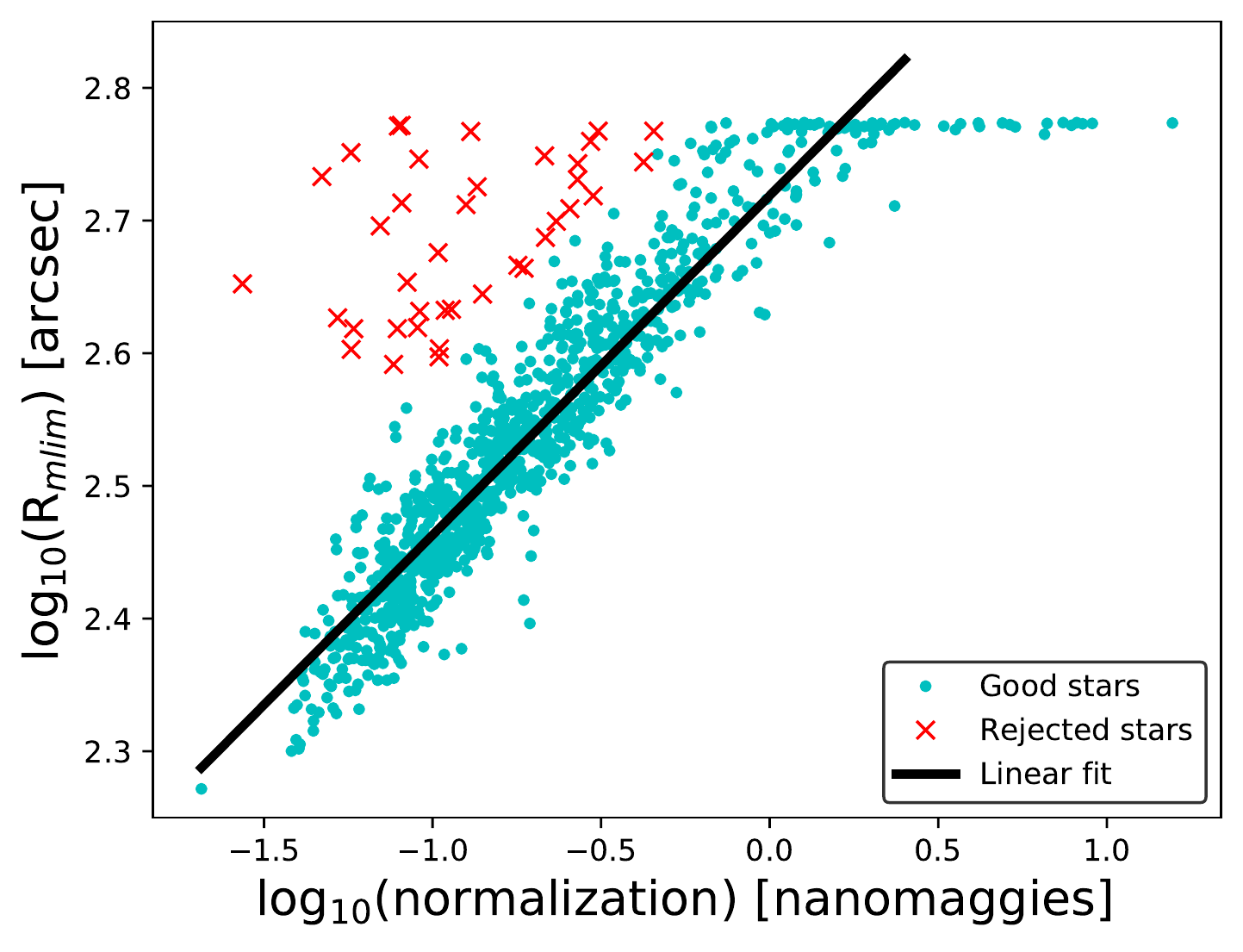}
    \caption{Removal of stars that are potentially affected by external (undesired) light in the outermost parts for the \textsl{g} SDSS band.
             We plot the radial distance ($\text{R}_{mlim}$) at which the surface brightness profile of each star reaches 26.5 mag arcsec$^{2}$ (the surface brightness limit of the SDSS survey) versus the normalization value estimated at the radial distance of 60 arcsec.
             As expected, the vast majority of the stars (blue points) follow a linear correlation trend between the two quantities, except for some cases, where the stars have an artificially large $\text{R}_{mlim}$ caused by parasitic light (red crosses).
             The outliers are identified through a 3$\sigma$ clipping mean taking the linear fit as reference (black solid line).
             The saturation of stars in the upper part of the plot with high-normalization values correspond to those stars which reach $\text{R}_{mlim}$ at the border of the image (which has the same size for all the individual frames).}
    \label{fig:psf_g_filtering}
\end{figure}

An additional advantage of estimating $\text{R}_{mlim}$ is that, combined with the normalization value we have used before to combine the stars, it can be used to remove the presence of stars heavily affected by contaminating sources.
Example of such contamination could be the presence of another close-by bright star, presence of Galactic cirri, very crowed region with many sources, etc.
The way we identify these potentially problematic cases is by plotting $\text{R}_{mlim}$ versus the normalization factor (see Fig.~\ref{fig:psf_g_filtering}).
In an ideal case (as all the stars used for creating the final stacked PSF have similar shapes), there should be a strong relation between the brightness at radial distance of 60 arcsec (i.e. the normalization factor we have used) and the radial distance where the brightness is 26.5 mag/arcsec$^2$ (i.e. $\text{R}_{mlim}$).
Figure~\ref{fig:psf_g_filtering} shows that this is in fact the case.
The vast majority of stars (blue points) follow a nice trend between these two quantities.
We can characterize this trend by fitting a linear relation (black solid line).
The 3$\sigma$ outliers to this relation are plotted using red crosses.
As expected, most of the outliers correspond to stars that have extra (parasitic) light from other nearby sources (that have not been properly masked), which artificially enlarge the location of $\text{R}_{mlim}$.
By removing those stars we create an accurate representation of the outer part of the PSFs.
The final stacking of all "good" (normalized and masked) stars is done through a 3$\sigma$ clipped median (pixel-by-pixel).
The removal of the outliers decrease our number of used stars to approximately 1000\footnote{The exact number of stars considered for the outer, intermediate, and inner parts can be found in the header of the PSF images with the keyword \texttt{NOUTER}, \texttt{NINTER}, and \texttt{NINNER}, respectively.}.

\subsection{Constructing the intermediate part of the PSFs}
\label{sec:method_psf_intermediate}
To build the intermediate part of the PSFs, we have followed essentially the same steps as the ones described in the previous Sect.~\ref{sec:method_psf_outer}.
This time we have used \nstars{} stars of magnitude \textsl{B1mag}$\sim$8.3 mag for \textsl{u} SDSS band, \textsl{R1mag}$\sim$9.0 mag for \textsl{g} and \textsl{r} SDSS bands, and \textsl{Imag}$\sim$8.3 mag for \textsl{i} and \textsl{z} SDSS bands.
Stars with this brightness have saturated pixels in the centre region.
However, the bleeding pixels have been well corrected by the SDSS pipeline, and they do not have the problem of the depletion explained in the previous section, and thus, no rejection of stars was done because of this issue.

The normalization of the stars for creating the intermediate part of the PSF is conducted in a ring of 1 pixel width at a radial distance from the centre of 7 arcsec.
As it was done for the outer part of the PSFs, two iterations are carried out in order to obtain better masks for undesired objects and also to improve the normalization of each star.
Finally,  we stacked all the masked and normalized stars using 3$\sigma$ clipping median to obtain the intermediate part of the PSF.

In this case, contrary to what happened for the outer part of the PSFs, all stars have the same brightness and therefore, their contribution in signal and noise to the final stacked image is very similar.
Because of that, it was not necessary to mask the outer regions of these stars according to their S/N.

\subsection{Constructing the central part of the PSFs}
\label{sec:method_psf_core}
To build the central part of the PSFs, \nstars{} stars with brightness $>$14 mag in all bands from USNO-B1 Catalog (\textsl{B1mag} for \textsl{u} SDSS band, \textsl{R1mag} for \textsl{g} and \textsl{r} SDSS bands, and \textsl{Imag} for \textsl{i} and \textsl{z} SDSS bands) were considered.
With this selection criteria, all the retrieved stars were unsaturated and therefore we can obtain the core of the PSFs by simply stacking them after a proper treatment.

The creation of the outer and intermediate parts of the PSFs, using the centre given by USNO-B1 Catalog (using low proper motion stars), was accurate enough for our purpose.
Indeed, due to the strong saturated and bleeding pixels, it was impossible to obtain the centre with more precision than the one given by the catalogue.
However, when constructing the central part of the PSF we can obtain the centre of each star more accurately, as we are considering non-saturated stars.
Note that this is important because small errors on the location of the central part create (after stacking many stars) wider PSFs than the true ones.
We use \texttt{Imfit} \citep{erwin_2015_Imfit} to fit a 2-dimensional Moffat function to each star \citep[see e.g.][]{trujillo_2001b}.
\texttt{Imfit} provides the spatial coordinates of the centre of the stars. Then we use \texttt{SWarp} to resample the star into a new grid with the centre obtained from \texttt{Imfit}.

In the case of the stars used to create the central part of the PSFs, \texttt{NoiseChisel} and \texttt{Segment} was run only once as the stars are faint enough, so a first detection and segmentation is able to properly mask all external sources.
The normalization of each star contributing to the stacked PSF is done in a ring of 1 pixel width at a radial distance of 2 arcsec.
Once all images are ready, we stack them using 3$\sigma$ clipping median to obtain a first version of the inner part of the PSF (psf1).

\begin{figure}
	\includegraphics[width=\columnwidth]{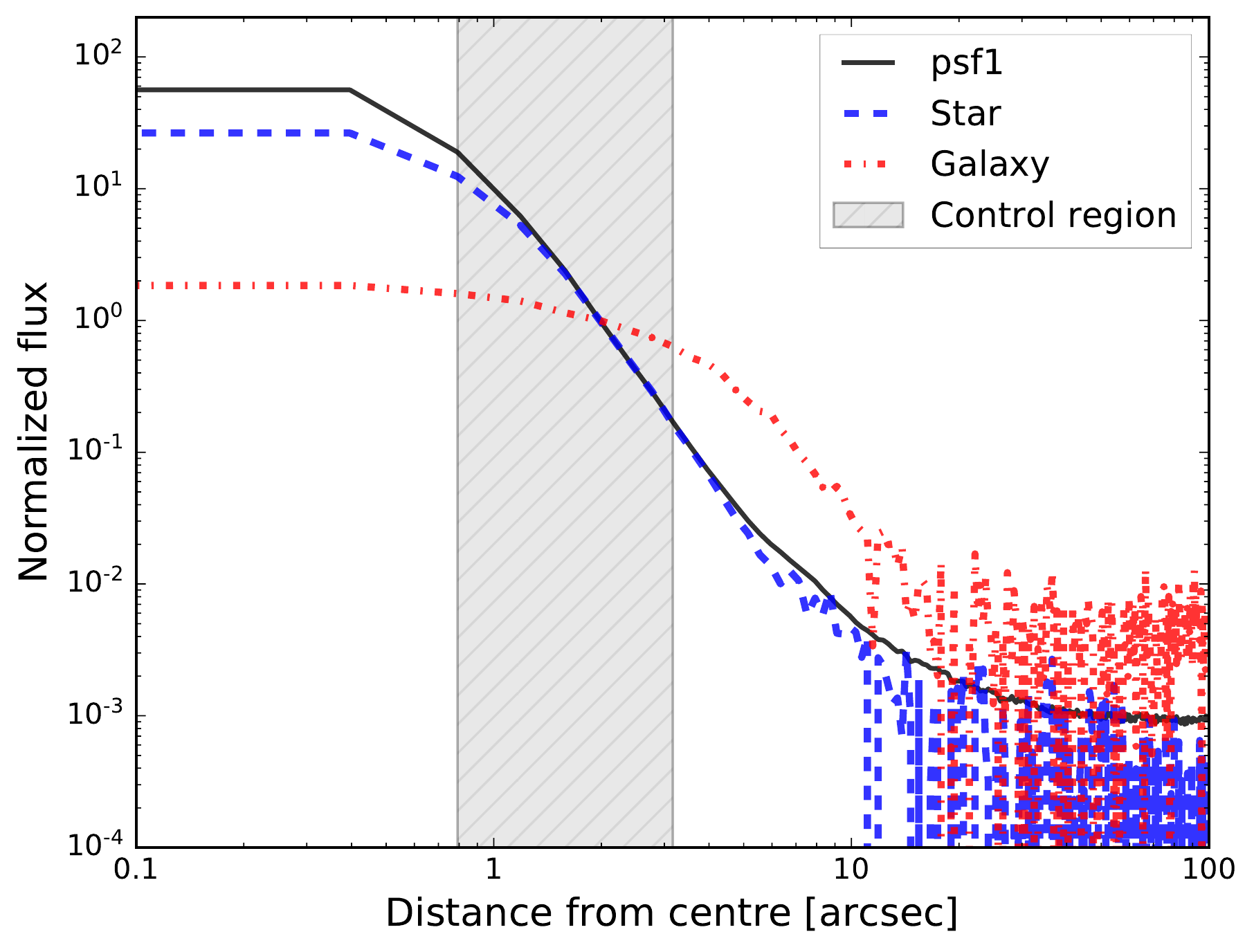}
    \caption{Radial profile of the stacked PSF (psf1; solid black line), an individual star (dashed blue line), and a contaminant galaxy (dot-dashed red line).
             The profiles are shown for \textsl{r} SDSS filter.
             The chosen star is at R.A. (J2000) = 16:13:49.55; Dec (J2000) = +32:21:27.43 while the galaxy is at R.A. (J2000) = 13:17:51.44; Dec (J2000) = +33:49:34.45.
             The grey region represents the radial location where the parameter $\text{s}$ (Eq.~\ref{eq:reject_core_stars}) is computed.
             All profiles intersect at a radial distance of 2 arcsec because it is the normalization radius.
             The parameter $\text{s}$ is used to distinguish between stars and extended objects.
             For this particular case, the values are $\text{s}=\svalueforstar$ (star) and $\text{s}=\svalueforgalaxy$ (galaxy).}
    \label{fig:plot_reject_core_profiles}
\end{figure}

An additional consideration that must be taken into account is the contamination of the USNO-B1 Catalog by sources that are not stars when selecting faint objects.
That problem was not an issue in the construction of the outer and intermediate parts of the PSFs.
This is because when considering very bright sources, the contaminants of the catalogue by sources that are not stars is negligible.
The main contaminant source of the catalogue at fainter magnitudes is produced by galaxies.
After obtaining the first version of the inner part of the PSF (psf1), we compare the stacked radial profile with the profiles of all the individual sources used to create psf1.
In Fig.~\ref{fig:plot_reject_core_profiles} we show the radial profile of the psf1 in solid black line, the radial profile of one bonafide star in dashed blue line, and the radial profile of one genuine galaxy in dot-dashed red line.
We use the fact that the radial profile of galaxies and stars are different in a region centred at the normalization radius (2 arcscec) to reject galaxies from our sample.
To remove the extended sources, we measure the following parameter:

\begin{equation}
    \text{s} = \text{std} \left( {\frac{\text{object}}{\text{psf1}}} \right)_{\text{Control area}}.
	\label{eq:reject_core_stars}
\end{equation}

\begin{figure}
	\includegraphics[width=\columnwidth]{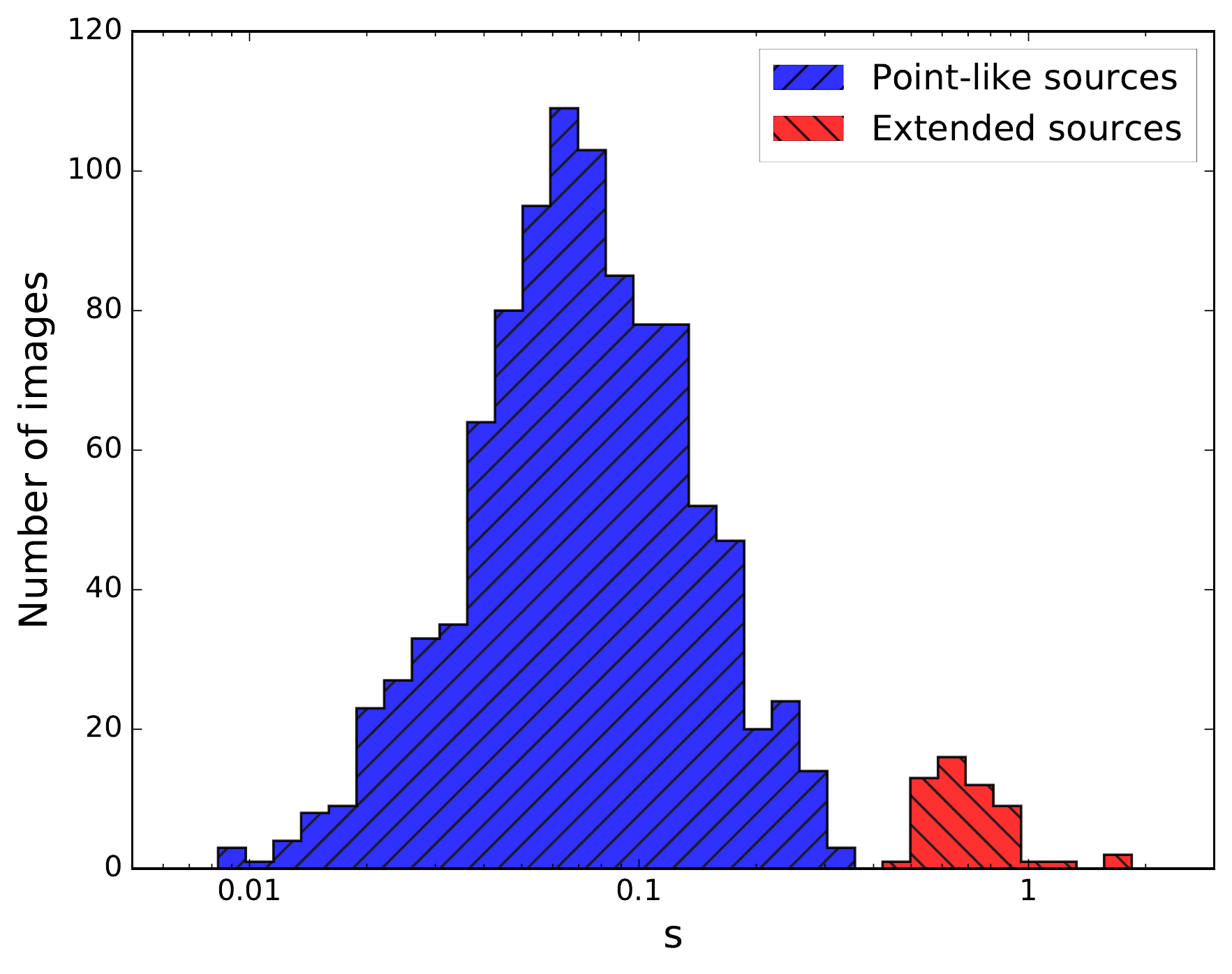}
    \caption{The removal of extended sources to create the central part of the PSFs.
             This figure shows the distribution of the parameter $\text{s}$ in the SDSS \textsl{r} filter.
             The vast majority of the sources has a small value of $\text{s}$ with a median $\text{s}$ value equal to \svaluefordip.
             Note that x-axis is in logarithmic scale.
             We have selected the dip between the blue and red distributions to separate our sample into bonafide stars and galaxy contaminants (see the text for details).}
    \label{fig:plot_reject_core_hist}
\end{figure}

Defined in this way, $\text{s}$ is the standard deviation of the ratio between the intensity radial profile of the object (star or galaxy) and the intensity radial profile of the psf1 in the "Control area" (shaded region in Fig.~\ref{fig:plot_reject_core_profiles}).
This region has 6 pixels width and is centred at the normalization radius (2 arcsec).
The "Control area" is chosen such that it maximizes the difference at measuring $\text{s}$ between the profile of a star and the profile of an extended source.
The radial profile of a genuine star will have a very similar shape compared to the radial profile of psf1, and therefore a low value of $\text{s}$ is expected.
On the other hand, the radial profile of a galaxy would differ significantly from the radial profile of psf1, and because of that the $\text{s}$ parameter will be larger.
The distribution of $s$ values is shown in Fig.~\ref{fig:plot_reject_core_hist} and is used to distinguish between stars and galaxies.
There are two clear peaks on the $\text{s}$ distribution.
The vast majority of objects are point-like sources and consequently, they are grouped around low values of $\text{s}$ ($\sim$0.07).
In addition to this, we find another group of extended objects with $\text{s}$ values 10 times higher.
These objects are removed from the final sample by using 2.5$\sigma$ clipped mean.
Interestingly, in the USNO-B1 Catalog there is a star-galaxy estimator that measures the probability of a particular source to be a star (point source like object) or galaxy (extended source).
However, we did not find an obvious correlation between the value of $\text{s}$ that we measure and the star-galaxy estimator in the USNO-B1 Catalogue.
Moreover, we found some sources classified as stars by USNO that, after visual inspection, turned out to be galaxies.
Because of that, in this work, we only rely in the parameter $\text{s}$ for rejecting extended sources from our sample.
The number of objects removed from the stacking due to this galaxy contamination depends on the filter, but on average the amount of objects rejected is about 20\% of the original sample (\nstars{} stars).
Once they were removed, we repeat the entire process of stacking the images using 3$\sigma$ clipped median to obtain the central part of the PSFs.

\subsection{PSF junction (central-intermediate-outer parts)}
\label{sec:method_psf_junction}
In this section we describe how the different parts of the PSF are joined together in order to produce a representative final PSF.
To conduct such process, we use the outer part of the PSF as the reference.
Then, the intermediate PSF is modified in such a way that its flux and reference sky level matches the shape of the outer PSF at a given radius, called the junction radius ($\text{r}_{jun}$).
Once this junction is done, the same process is done for joining the outer-intermediate PSF with the central PSF\footnote{The exact junction radius for joining the outer and the intermediate parts can be found in the header of the images with the keyword \texttt{ROUTER}. For the junction radius of the outer-intermediate with the central part, the keyword is \texttt{RINNER}.}.

\begin{figure}
	\includegraphics[width=\columnwidth]{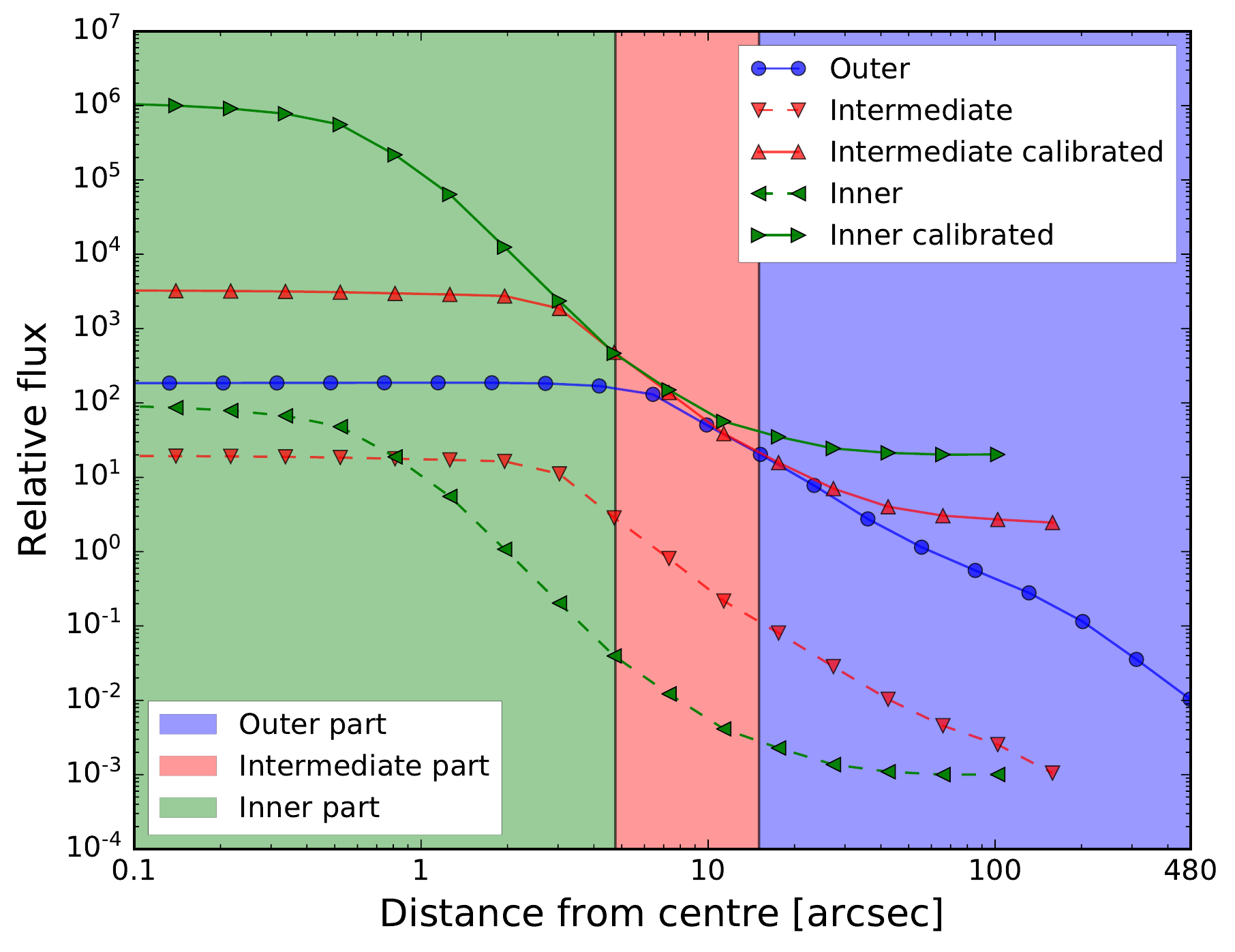}
    \caption{Radial profiles of the different PSF parts in the case of the SDSS \textsl{r} band.
             Dashed lines are radial profiles before the junction and solid lines are radial profiles after the matching.
             The radial profile of the outer PSF part (solid line with circles) is taken as the reference.
             The different shaded regions are colour coded in the bottom legend and correspond to the three pieces of the PSF.
             The vertical lines that divide the different coloured regions indicate the radius at which these parts have been joined.}
    \label{fig:plot_psf_junction}
\end{figure}

The region in which the junction is done depends on each SDSS band, but the criteria is the same for all bands.
We require equal S/N between the different PSF regions at the junction radius in order to have a smooth transition between the different parts.
Therefore, the choice of $\text{r}_{jun}$ is calculated from the S/N radial profiles of the different sections of the PSF.
Figure~\ref{fig:plot_psf_junction} shows the joining methodology for the PSF in the SDSS \textsl{r} band.
As commented above, the PSF radial profile of the outer part is used as the reference (solid line with circles).
For the \textsl{r} band, the saturation of the brightest stars appears at a radial distance of $\sim$10 arcsec.
This can be seen as a flat radial profile for radii smaller than 10 arcsec.
The radial profile of the intermediate part is plotted using a dashed line with down oriented triangles.
This intermediate part has a different flux compared to the outer PSF part.
The junction of these two PSF pieces consists in modifying the intermediate PSF to have the same flux (a multiplicative factor $\text{f}$) and reference sky background level (additive factor $\text{c}$) than the profile of the outer PSF.
Having two parameters to determine ($\text{f}$ and $\text{c}$) we need two independent equations to solve the problem.
We do the following, we use the junction radius to separate two different regions in the profiles.
A region located at a radius $\text{r}_{-1}$ (which is $\text{r}_{jun}-1$ pixel) and another region located at $\text{r}_{+1}$ (which is $\text{r}_{jun}+1$ pixel).
The equations we solve are

\begin{equation}
    \label{eq:eq_renormalization1}
    \begin{aligned}
    \text{O}_{-1} &= \text{I}_{-1} \times \text{f} + \text{c} \\
    \text{O}_{+1} &= \text{I}_{+1} \times \text{f} + \text{c}
    \end{aligned}
\end{equation}

where $\text{O}_{-1}$ and $\text{O}_{+1}$ are the values of the radial profile of the outer PSF part at $\text{r}_{-1}$ and $\text{r}_{+1}$, respectively.
$\text{I}_{-1}$ and $\text{I}_{+1}$ are the values of the radial profile of the intermediate PSF part at $\text{r}_{-1}$ and $\text{r}_{+1}$, respectively.
Consequently, $\text{f}$ and $\text{c}$ are

\begin{equation}
    \label{eq:eq_renormalization2}
    \begin{aligned}
    \text{f} &= \frac{\text{O}_{+1} - \text{O}_{-1}}{\text{I}_{+1} - \text{I}_{-1}} \\
    \text{c} &= \text{O}_{-1} - \text{I}_{-1} \times \frac{\text{O}_{+1} - \text{O}_{-1}}{\text{I}_{+1} - \text{I}_{-1}}
    \end{aligned}
\end{equation}

The junction between the central PSF part (dashed line with right oriented triangles in Fig.~\ref{fig:plot_psf_junction}) and the matched outer-intermediate PSF is done in the same way as in the case of the outer and intermediate PSFs junction.
Note that the plateau region observed in the most inner radial profile (solid line with left oriented triangles) is not saturation but the value of the central pixel of the PSF.
Considering these three PSF parts, we are able to obtain the entire extended PSFs without saturation in the inner part and high S/N in the outer part.

\subsection{Final refinements to the extended PSFs}
\label{sec:final_considerations}
The outer parts of the PSFs have been built using very bright stars.
As a consequence, the final PSF still contains a rectangular region along the central part of the PSF and perpendicular to the drift scanning direction that presents a depletion.
This is an artificial effect due to the saturation of the brightest stars and remains after all the corrections made by the SDSS pipeline.
To correct for this effect in our final PSFs, we interpolate the values of the pixels in these depleted regions using the flux values of adjacent (non-saturated) pixels.

Finally, the PSFs are rotated 90 deg counterclockwise to orientate the drift scanning direction with the left to right direction.
That means that in our final PSFs the sky would move from left to right.
To use our PSFs, they must be rotated to match the drift scanning direction of the given SDSS data set.
In the particular case of the Stripe82 survey, since it observes an equatorial band, the sky is moving from left to right and consequently, this is the same direction than the orientation of the final PSFs.
In any other case, the PSFs will have to be rotated to align the drift scanning direction of the image with the drift scanning direction of the PSF.
To finish the process, we normalize each PSF to have the sum of all pixels equal to 1.

\section{Results}
\label{sec:results}
By following the previous steps it has been possible to obtain PSFs in the five SDSS bands with an unprecedented radial extension ($\sim$8 arcmin).
All PSFs obtained in this work are publicly available and they can be found at the IAC Stripe82 web page in FITS format\footnote{Version of PSFs described in the current paper is \texttt{\projectversion}. The version number of each PSF is stored into the header of the images and can be checked with the keyword \texttt{VERSION}.}: \href{http://research.iac.es/proyecto/stripe82/pages/advanced-data-products/the-sdss-extended-psfs.php}{http://research.iac.es/proyecto/stripe82/pages/advanced-data-products/the-sdss-extended-psfs.php}.
It is important to bear in mind that PSFs provided above have to be rotated in order to align the drift scanning direction with the drift scanning direction of the image considered.
Section~\ref{sec:results} is divided in two parts. Section~\ref{sec:results_characteristics} is dedicated to describe the characteristics and features of the extended PSFs.
In Sect.~\ref{sec:results_scatterlight}, we use the PSFs to model and remove the scattered light field of the stars in the SDSS image of the Coma Cluster central region as a practical example.

\subsection{PSFs characteristics}
\label{sec:results_characteristics}
\begin{figure*}
	\includegraphics[width=\textwidth]{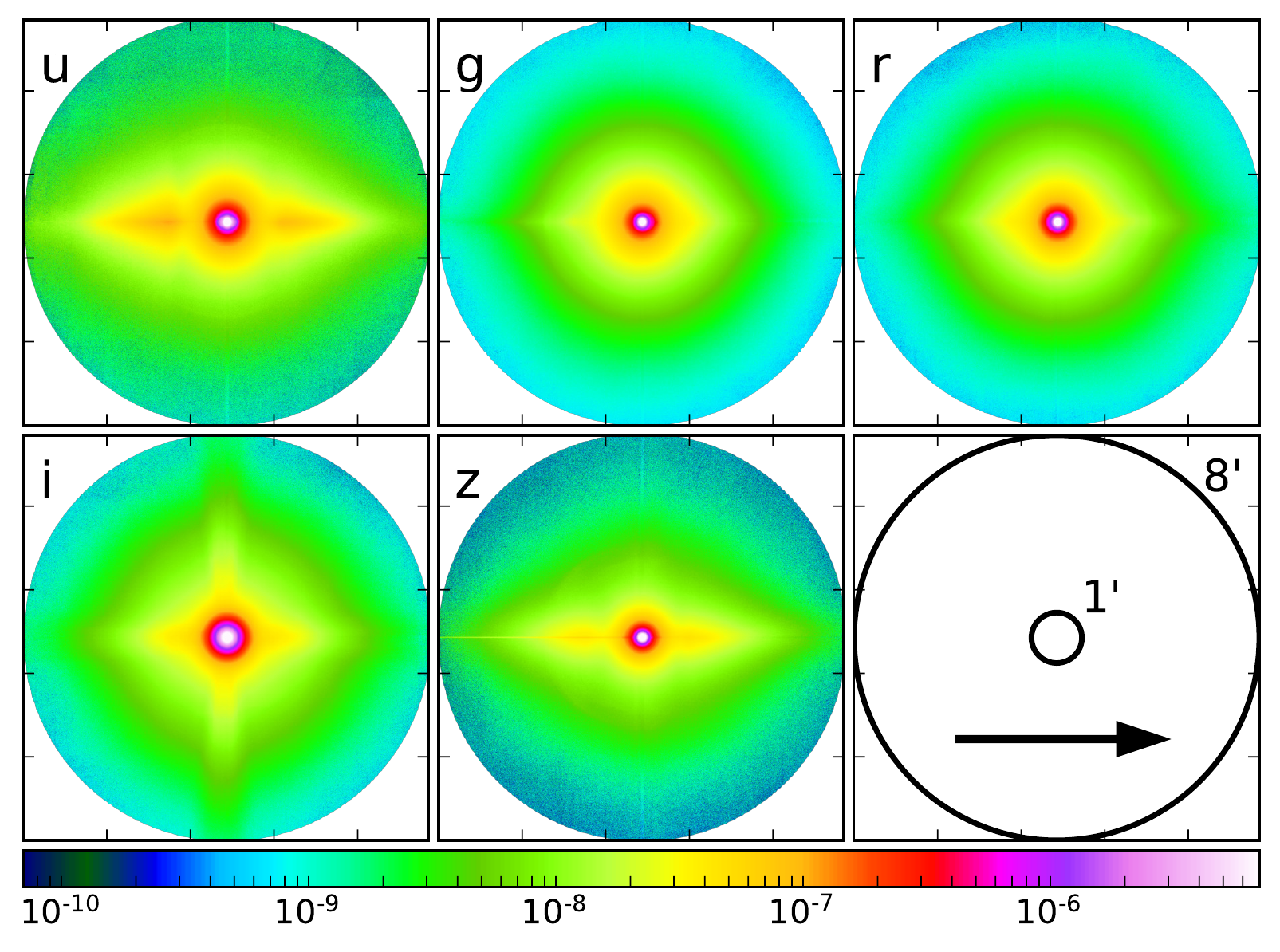}
    \caption{The extended SDSS PSFs.
             The bottom right panel illustrates the size of the PSFs showing, with two circles, the location of the radial distances of 1 and 8 arcmin.
             The arrow orientation indicates the drift scanning direction.
             All PSFs have been normalized to have a total flux of 1.
             The colour bar indicates the pixel values of each PSF.
             Note that the intensity range of the pixel extends near 5 orders of magnitude.}
    \label{fig:plot_psfs_2d}
\end{figure*}

The extended SDSS PSFs can be seen in Fig.~\ref{fig:plot_psfs_2d}.
Along the entire structure, the S/N radial profiles of the \textsl{g}, \textsl{r}, and \textsl{i} PSFs attain values above 3 while the \textsl{u} and \textsl{z} PSFs have S/N$\sim$1 at a radial distance of 8 arcmin.
Beyond that radius, the quality of the PSFs drops significantly.
This is because at larger distances, the low number of images used to create the PSFs are manifested through the appearance of residuals left by the object masks.
Because of that, we have cropped the PSFs to have an extension in radius of 8 arcmin for all SDSS bands.
All pixels beyond that distance from the centre (8 arcmin) have been masked with \texttt{NaN} (Not a Number) values.
In any case, depending on the user's need, it is possible to crop the PSFs
to a smaller size but taking into account that the normalization of the extended PSFs are done such that the sum of all (not \texttt{NaN}) pixels is equal to 1.

It is important to note that the structural properties of the PSF vary depending on multiple parameters, such as the camera column number, seeing, spatial location on the CCD, and others \citep[see][for further details about the variation of SDSS PSFs]{gunn_2006_sdss_telescope,xin_2018}.
In this sense, the extended PSFs we present in this work represent the properties of averaged PSFs for the entire SDSS survey.

Although we have done our best during the masking process, including a double iteration in order to avoid internal reflections (i.e. the reflections of light produced after reflecting of the CCD and reflecting again in optical elements such as the filter, etc.), some residuals remain in the final PSFs. In each band they appear at different positions.
For example, in the case of \textsl{g} and \textsl{r} PSFs, the internal reflection appears at a radial distance of $\sim$25 arcsec from the centre. In the case of the \textsl{i} band PSF, the internal reflections are located in the centre.
The fact that the internal reflections of \textsl{g}, \textsl{r}, and \textsl{i} filters are near to their centres make it possible to subtract the scattered light produced by the stars without leaving behind strong residuals (we will show a practical example in the next section).
In the case of the \textsl{u} and \textsl{z} bands, the PSFs present two big lobes of light in the left and right sides.
These structures are aligned parallel to the drift scanning direction and they vary from exposure to exposure.
It is therefore challenging to subtract the light of stars in \textsl{u} and \textsl{z} filters without producing strong residuals, although these residuals are stronger in the \textsl{u} band.

The diffraction spikes, produced by the secondary mirror structure of the telescope that are visible in individual images of stars in the SDSS survey, do not appear in our combined PSF models.
This is because of the random orientation of the spikes in individual exposures.
For this reason, they are averaged out in the final stacked PSFs.
Additionally, in order to increase the S/N of the final models, the spikes were masked in the second masking iteration.
We originally tried to include the spikes in our final PSFs, but the variability in flux and shape made impossible to model them realistically in the end.

\subsection*{The effect of the drift scanning in the PSFs}
\label{sec:driftscann_effect}
\begin{figure}
	\includegraphics[width=\columnwidth]{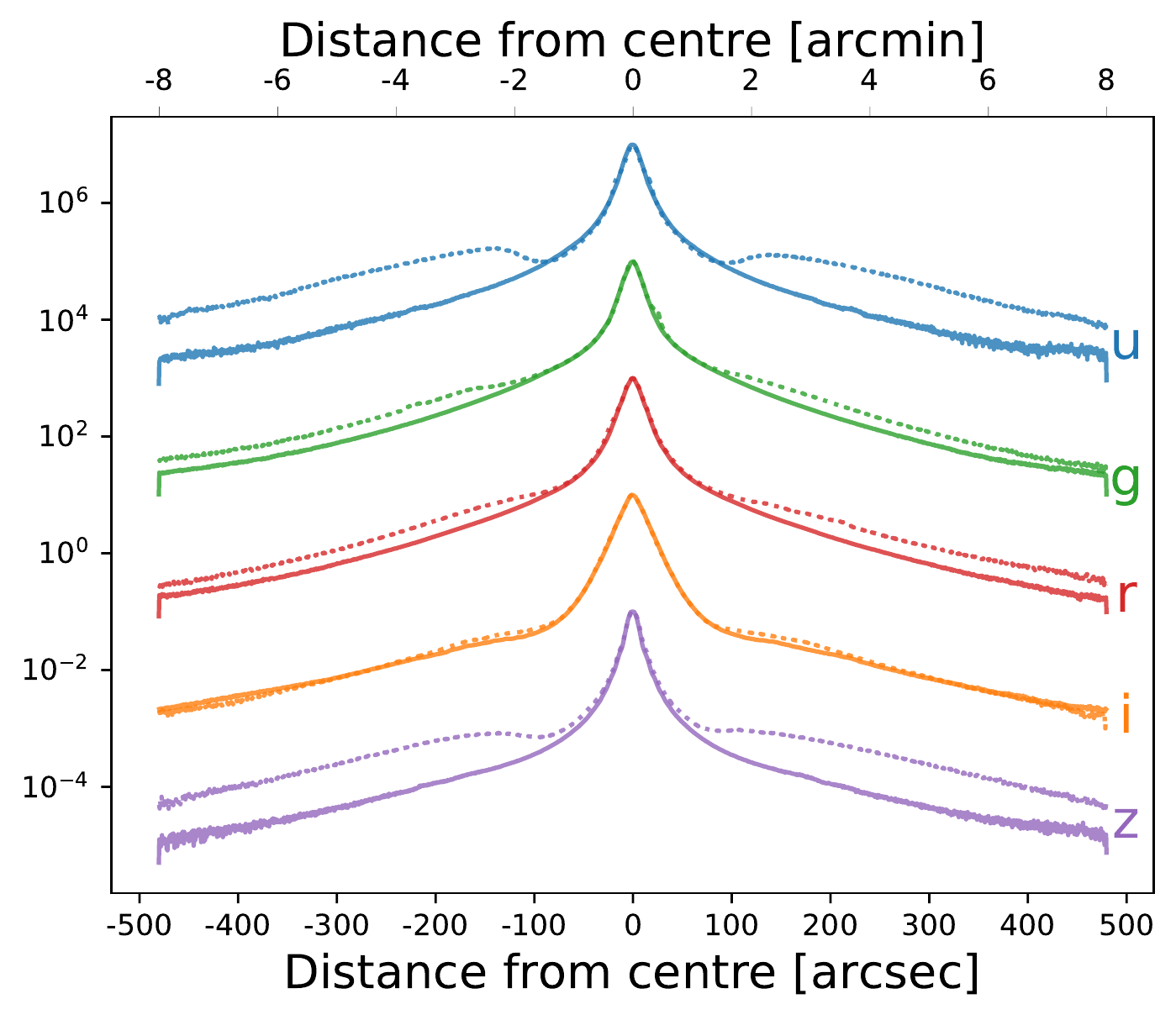}
    \caption{PSF profiles along the drift scanning direction (dotted lines) and perpendicular to it (solid lines).
             Each SDSS filter is annotated on the right side of the figure.
             To ease the reading, PSF profiles have been shifted vertically.
             The vertical axis is shown in arbitrary units.
             The horizontal axis corresponds to the radial distance from the centre in arcsec (bottom) and in arcmin (top).}
    \label{fig:plot_driftscanning}
\end{figure}

The drift scanning observing mode of the SDSS survey makes the PSFs somewhat asymmetric as is seen in Fig.~\ref{fig:plot_psfs_2d}.
In order to quantify this asymmetry, we study the flux along the drift scanning direction and along the perpendicular direction.
We have done that by measuring the flux along these two directions using slits 50 pixels wide.
As is seen in Fig.~\ref{fig:plot_driftscanning}, the \textsl{g} and \textsl{r}-band PSFs have an excess of flux in the drift scanning direction beyond 2 arcmin of radial distance from the centre (dotted lines compared with solid lines).
This excess of flux is quantified by computing the ratio between the two profiles and the results are that along the drift scanning direction the flux is \gfluxexcess{} and \rfluxexcess{} times higher than in the perpendicular direction, respectively.
The \textsl{i} band PSF presents a cross-like shape structure whose main arms correspond to the drift scanning and perpendicular directions.
In the case of \textsl{u} and \textsl{z}, PSFs there are two bright lobes along the drift scanning direction.
Along such direction, the fluxes are \ufluxexcess{} and \zfluxexcess{} times higher than in the perpendicular direction, respectively.

All these features demonstrate that an extrapolation of the PSFs to larger radii assuming a power law or any other particular function is too crude an approximation to represent the structural richness of the SDSS PSFs.
Because of the non-symmetric shape of the PSFs, it is also necessary to rotate the PSF according to the orientation of each particular field when PSF models are used to subtract stars (see Sect.~\ref{sec:results_scatterlight}).
The same is also necessary when the PSFs are used to model the effect of the PSF in astronomical objects with the aim of obtaining their intrinsic physical parameters (e.g. when using \texttt{Imfit}, \texttt{GALFIT}, or similar astronomical software).

\subsection*{Radial profiles of SDSS PSFs}
\label{sec:radial_profiles}
\begin{figure*}
	\includegraphics[width=\textwidth]{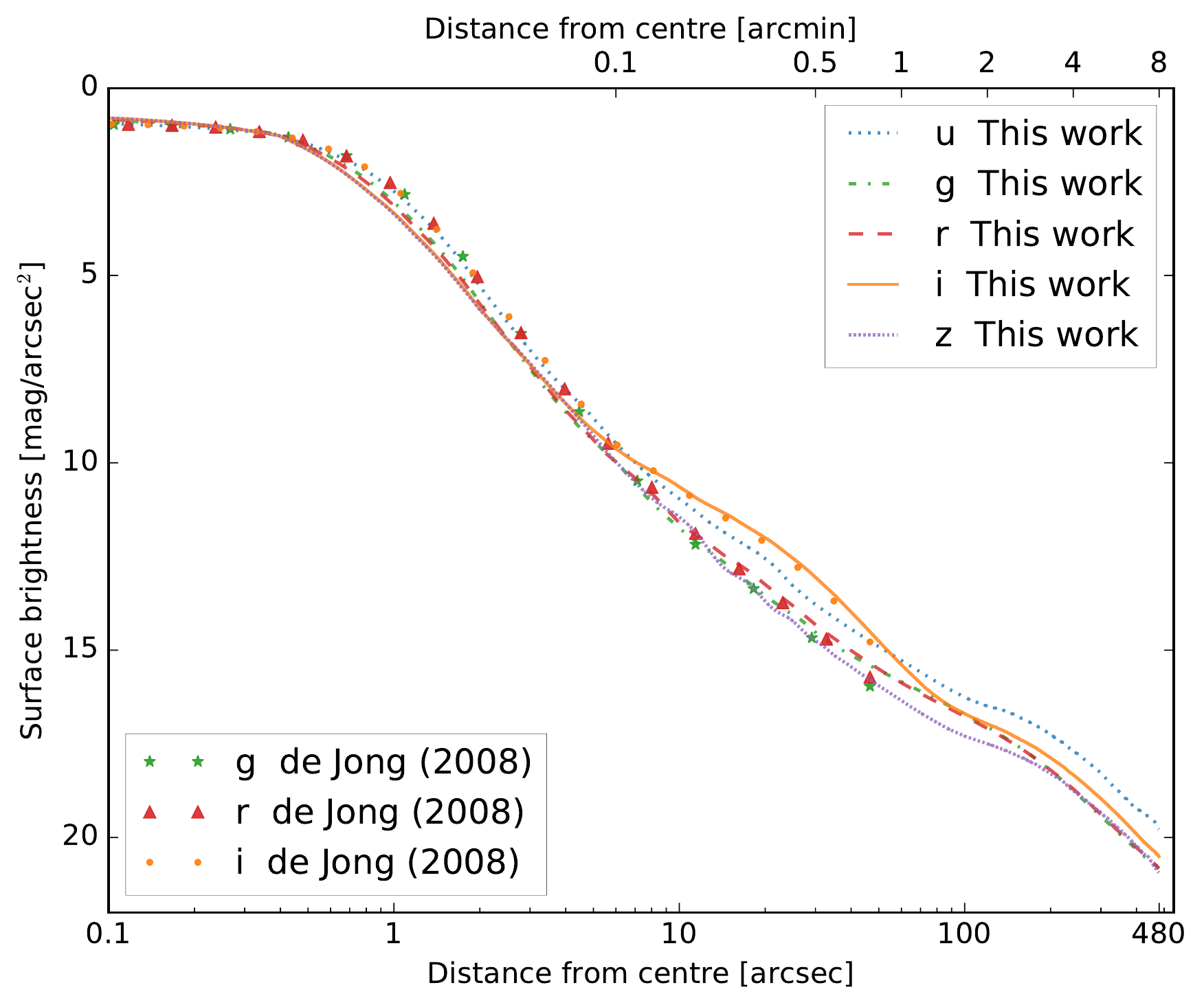}
    \caption{Circulary averaged surface brightness radial profiles of SDSS PSFs of this work compared with those obtained by \protect\cite{dejong_2008}.
             The SDSS filters are line coded as explained in the legends.
             A remarkable feature is the bump in the \textsl{i} band that can be seen as an extended halo of light between 7 and 70 arcsec.}
    \label{fig:plot_profiles1d}
\end{figure*}

Figure~\ref{fig:plot_profiles1d} shows the circulary averaged radial profiles of all SDSS PSFs.
For comparison, we also plot the radial profiles of the SDSS PSFs obtained by \cite{dejong_2008}.
To facilitate the comparison of the two works, we have re-normalized the flux of our PSFs to 1 only accounting for the flux inside the maximum radius of the profiles produced by \cite{dejong_2008} (i.e. $\text{R}_{max}\sim46$ arcsec).
The dynamical range of our PSFs covers 20 mag, and the PSF radial profiles decline as power laws ($r^{-\alpha}$) with an exponent close to $\alpha=\meanslopevalue$

\subsection{Scattered light subtraction}
\label{sec:results_scatterlight}

In this section we show an example of how the PSF models can be used to remove the scattered light produced by the stars in the central part of the Coma Cluster field.
We select this region because it is a classic example where studying intracluster light or low-surface brightness structures in general is impossible without an extended and well-characterized PSF.

We have removed the scattered light field produced by all the stars in the Gaia DR2 catalogue \citep{gaia2_2018} with $\text{Gmag}<13$ mag.
We only show this exercise using \textsl{g}, \textsl{r}, \textsl{i} filters.
These images are deeper compared with the other two filters (\textsl{u} and \textsl{z}) in the SDSS survey, and the effect of scattered light towards the faintest regions is therefore easier to show.
Note that in Sect.~\ref{sec:method_psf_outer} we used USNO-B1 Catalog, because we found Gaia was incomplete among the brightest stars.
In this particular field, the brightest star has $\text{Gmag}=7.0$ mag and it is part of the Gaia DR2 Catalogue.
Therefore, we used the more accurate star positions provided by Gaia.

Once a list of stars is ready, the scattered light field is modelled in a hierarchical way, that is, we start by fitting and subtracting the brightest star, then we go to the second brightest star, and so on.
The process is repeated until all stars of a given magnitude are fitted and subtracted. The reason of doing this iterative process is because the light of very bright stars affect significantly the fitting and subtraction of other near and fainter stars.
The fitting of each star requires two steps: first, to obtain an accurate centre of the star and secondly, to perform the flux calibration of the PSF.
The details of how the process works are as follow.

\begin{figure}
	\includegraphics[width=\columnwidth]{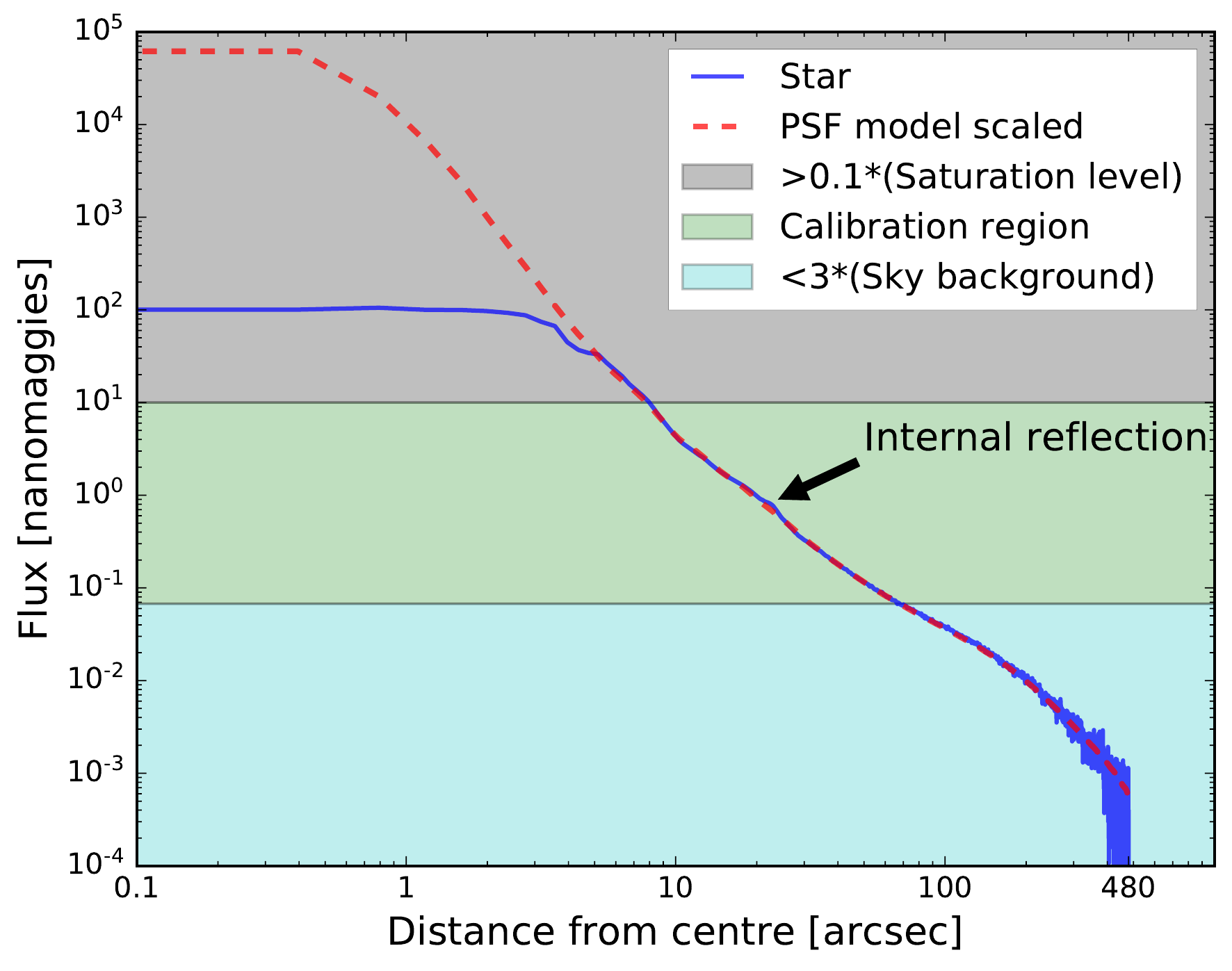}
    \caption{An example of PSF model fitting to a real star (R.A. (J2000) = 12:59:32.975, Dec (J2000) = +28:03:56.295, $\text{Gmag}=7.0$ mag) in SDSS \textsl{r} band.
             Solid blue line is the radial profile of the star to be modelled.
             The grey region corresponds to the radial profile values which are 0.1 times above the saturation level of the image ($\text{saturation=100}$ nanomaggies).
             The light blue area represents the flux level which are 3 times below the intensity value of the background value of the image.
             The light green area corresponds to the region used to obtain the scaling factor between the PSF model and the real star.
             The dashed red line represents the radial profile of the PSF once it has been scaled up to the same flux level as the real star.
             See the text for more details.}
    \label{fig:plot_star_fitting}
\end{figure}

First, we run \texttt{NoiseChisel} and \texttt{Segment} on the image to get the object masks as well as the background value and the saturation level.
As a first approximation, the centre of the stars to be modelled are obtained from the Gaia DR2 catalogue.
We refined the centre estimation by using \texttt{Imfit} (as explained in Sect.~\ref{sec:method_psf_core}).
The calibration in flux of the PSF model to each individual star is done through the radial profile of the star.
To calibrate the PSF, we use the radial range between 0.1 times the saturation level and 3 times the value of the background level of the image (see Fig.~\ref{fig:plot_star_fitting}).
In this region, for each pixel we compute the ratio between the radial profile of the star and the radial profile of the PSF.
The exact factor $\text{F}$ is computed by taking the 3$\sigma$ clipped median of all the pixels ratio.
$\text{F}$ is the factor by which the PSF model has to be scaled up to get the same flux level as the star.
Once we have the centre position of the star and F, the star is completely modelled, and we subtract the model from the original image.
As we have said above, this process is done star by star, until the last star considered is modelled and subtracted.

For modelling the scattered light field, a good masking of the sources in the image is mandatory.
Otherwise, the excess of flux from objects close to the stars that are being modelled could alter the measurement of their centre and flux.
To minimize these problems, we perform the scattered light modelling in two iterations.
The first one is the one that has been described above.
Once we have removed the brightest stars from the image, we run again \texttt{NoiseChisel} and \texttt{Segment} over the image to obtain a better object masks.
By doing this, the masking process takes into account the diffraction spikes, internal reflections and nearby objects that were not possible to obtain in the first masking.
Finally, we repeat again the scattered light field fitting process taking into account the new object masks.
Both the centre position and the scaling factor $\text{F}$ are better determined with the improved masks in the second iteration.
Note that this two-step methodology is the same as described in Sect.~\ref{sec:method_psf_outer} for constructing the outer and intermediate part of the PSFs, but in this case for subtracting the stars.

\begin{figure*}
	\includegraphics[width=\textwidth,height=0.91\textheight,keepaspectratio]{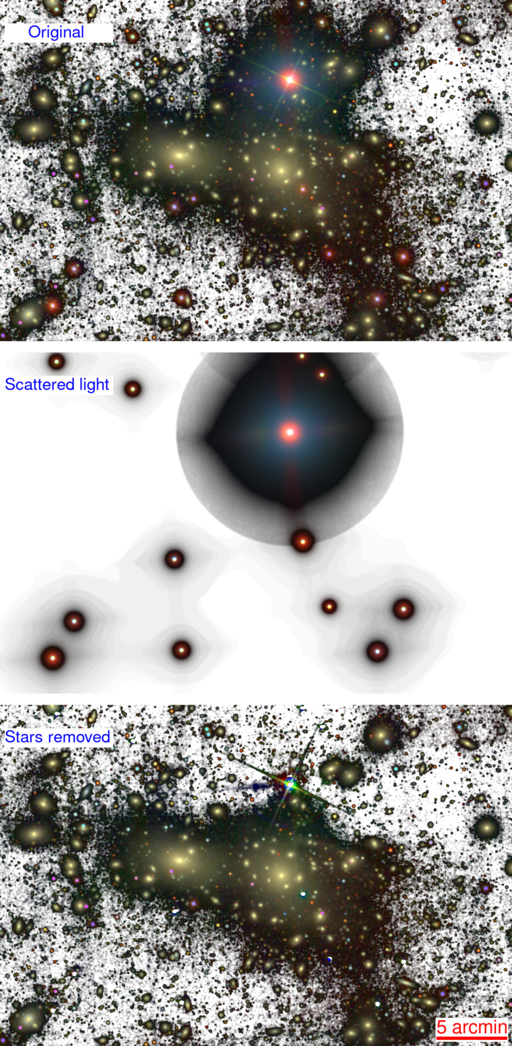}
    \caption{Colour composite image of the central region of the Coma Cluster as seen by SDSS (R.A. (J2000) = 12:59:48.658, Dec (J2000) = +27:58:49.443), using \textsl{g}, \textsl{r}, and \textsl{i} filters.
             The background of the image (in grey) corresponds to the sum of \textsl{g}, \textsl{r}, and \textsl{i} SDSS band images.
             The images have been rebinned to 2 arcsec per pixel and convoluted with a gaussian kernel of 3 pixel width to enhance the low-surface brightness features of the data.
             Upper panel: SDSS original image.
             Note the enormous contamination produced by the brightest stars in the central part of the Coma Cluster.
             Middle panel: scattered light field of all stars brighter than 13 mag in the \textsl{G} band of Gaia DR2 Catalogue.
             Bottom panel: scattered light field subtracted image.}
    \label{fig:plot_coma_panel}
\end{figure*}

In Fig.~\ref{fig:plot_star_fitting} we show an example of how the PSF modelling process works for a real star.
The radial profile of a real star in the \textsl{r} band SDSS Coma Cluster central field is plotted in solid blue line.
The innermost region of the star is clearly saturated as can be seen from the flatness of the radial profile within $\text{R}<5$ arcsec.
A small bump at $\text{R}\sim25$ arcscec that corresponds to the internal reflection of the star is also seen.
The model PSF does not have this feature, because we masked them out during the second iteration of the PSF construction process.
As our PSF model cannot model all the particular features of a given individual star, after the subtraction process, some residuals are left.
They correspond mainly to the central region due to the saturated pixels and the internal reflections.
In the same way, the diffraction spikes of each star will remain as residuals since the PSF models do not have diffraction spikes.
All of these effects can be seen in Fig.~\ref{fig:plot_coma_panel}, where we show a colour composite (\textsl{g}, \textsl{r}, and \textsl{i}) image of the central part of the Coma Cluster field in three different panels.
Top panel shows the original image with the scattered light of near and bright stars contaminating the central parts of the galaxy cluster.
In the middle panel, we show the scattered light field model of the brightest stars.
In this particular case, we have modelled all stars brighter than 13 mag in the \textsl{G} band of Gaia DR2 Catalogue.
The bottom panel shows the stars subtracted image.
After the removal of the scattered light field, the low-surface brightness features of the Coma Cluster and its intracluster light are ready to be explored.

\section{Discussion and conclusions}
\label{sec:conclusions}
We present extended SDSS PSFs (8 arcmin in radius) for all five SDSS filters.
We describe the  methodology we have used to obtain these large PSFs and we show a particular example of how the PSFs can be used to remove the scattered light field produced by the brightest stars in the images.
Our approach of deriving PSFs can be applied to other telescopes and instruments to obtain extended PSFs.
This work shows that the PSFs have complex bidimensional structures as it can be seen in Fig.~\ref{fig:plot_psfs_2d}.
In that figure we plot the radial profiles produced by \cite{dejong_2008} with a maximum radius of $\sim$46 arcsec for the \textsl{g}, \textsl{r}, and \textsl{i} SDSS filters together with the radial profiles of the PSFs of this work.
The reader can see that the agreement between these two works is rather good.
Compared to \cite{dejong_2008}, we enlarge the model of SDSS PSFs an order of magnitude in radius.
In addition to this, we also provide their 2-dimensional structures.

The PSFs models presented in this work can be used to remove the scattered light of point-like sources, as we have described in Sect.~\ref{sec:results_scatterlight}.
In addition to this, the PSFs can be also used to model galaxies and other astronomical objects with the goal of correcting the scattered light of the object itself \citep[see e.g.][]{martinezlombilla_2019}.

The next generation of large and deep sky surveys such as LSST, Euclid mission, James Webb Space Telescope (JWST), and many others would require an exquisite characterization of their PSFs for studies involving low surface-brightness structures.
In future works, we will also show how scattered light produced by the extended objects themselves (in addition to that produced by point-like sources) is affecting their outskirts.

\subsection*{Note on reproducibility}
\label{sec:reproducibility}
Being able to reproduce results is in the core of science, and because of that, we have developed the construction of the PSFs in a reproducible way.
This means that the construction of the PSFs can be done by anyone from scratch, just downloading the project publicly available at \url{https://gitlab.com/infantesainz/sdss-extended-psfs-paper/} and following the instructions described in the \texttt{README.md} file of that repository.
All the results shown in this paper correspond to the Git commit \texttt{\projectversion}.
Note that the subtraction of the scattered light described in Sect.~\ref{sec:results_scatterlight} is not included, because it is not part of the PSF construction but a particular example on how to remove the scattered light produced by the stars in a given region of the sky.

\section{Acknowledgements}
We thank the referee for a constructive report that helped to improve the presentation of the manuscript.
This research has been supported by the Spanish Ministry of Economy and Competitiveness (MINECO) under grants AYA2016-77237-C3-1-P and AYA2016-76219-P.
We also acknowledge support from the Fundaci\'on BBVA under its 2017 programme of assistance to scientific research groups, for the project "Using machine-learning techniques to drag galaxies from the noise in deep imaging".
This work was partly done using the \href{http://savannah.nongnu.org/projects/reproduce/}{reproducible template project} (Akhlaghi et al. in prep).
The reproducible template was also supported by European Union’s Horizon 2020 (H2020) research and innovation programme via the RDA EU 4.0 project (ref. GA no. 777388).
We thank Mohammad Akhlaghi for all his time spent in explaining how to make the core part of this project reproducible.
We thank Roelof de Jong for kindly providing us the PSF profiles obtained in his work.
We thank Alejandro Borlaff, Nushkia Chamba, and Sim\'on D\'iaz-Garc\'ia for their comments.

This work was partly done using GNU Astronomy Utilities (\texttt{Gnuastro}, ascl.net/1801.009) version 0.10.
Work on \texttt{Gnuastro} has been funded by the Japanese Ministry of Education, Culture, Sports, Science, and Technology (MEXT) scholarship and its Grant-in-Aid for Scientific Research (21244012, 24253003), the European Research Council (ERC) advanced grant 339659-MUSICOS, European Unions Horizon 2020 research and innovation programme under Marie Sklodowska-Curie grant agreement No 721463 to the SUNDIAL ITN.

Funding for the Sloan Digital Sky Survey IV has been provided by the Alfred P. Sloan Foundation, the U.S. Department of Energy Office of Science, and the Participating Institutions.
SDSS-IV acknowledges support and resources from the Center for High-Performance Computing at the University of Utah.
The SDSS web site is www.sdss.org. SDSS-IV is managed by the Astrophysical Research Consortium for the Participating Institutions of the SDSS Collaboration including the Brazilian Participation Group, the Carnegie Institution for Science, Carnegie Mellon University, the Chilean Participation Group, the French Participation Group, Harvard-Smithsonian Center for Astrophysics, Instituto de Astrof\'isica de Canarias, The Johns Hopkins University, Kavli Institute for the Physics and Mathematics of the Universe (IPMU) / University of Tokyo, the Korean Participation Group, Lawrence Berkeley National Laboratory, Leibniz Institut f\"ur Astrophysik Potsdam (AIP), Max-Planck-Institut f\"ur Astronomie (MPIA Heidelberg), Max-Planck-Institut f\"ur Astrophysik (MPA Garching), Max-Planck-Institut f\"ur Extraterrestrische Physik (MPE), National Astronomical Observatories of China, New Mexico State University, New York University, University of Notre Dame, Observat\'ario Nacional / MCTI, The Ohio State University, Pennsylvania State University, Shanghai Astronomical Observatory, United Kingdom Participation Group, Universidad Nacional Aut\'onoma de M\'exico, University of Arizona, University of Colorado Boulder, University of Oxford, University of Portsmouth, University of Utah, University of Virginia, University of Washington, University of Wisconsin, Vanderbilt University, and Yale University.




\bibliographystyle{mnras}
\bibliography{paper.bbl}

\appendix

\section{Software acknowledgement}
This research was done with the following free
software programs and libraries: Bzip2 1.0.6, CFITSIO 3.47, CMake 3.15.3, cURL 7.65.3, Discoteq flock 0.2.3, FFTW 3.3.8 \citep{fftw}, File 5.36, FreeType 2.9, Git 2.23.0, GNU Astronomy Utilities 0.10 \citep{gnuastro}, GNU AWK 5.0.1, GNU Bash 5.0.11, GNU Binutils 2.32, GNU Compiler Collection (GCC) 9.2.0, GNU Coreutils 8.31, GNU Diffutils 3.7, GNU Findutils 4.7.0, GNU Grep 3.3, GNU Gzip 1.10, GNU Integer Set Library 0.18, GNU libiconv 1.16, GNU Libtool 2.4.6, GNU M4 1.4.18, GNU Make 4.2.90, GNU Multiple Precision Arithmetic Library 6.1.2, GNU Multiple Precision Complex library, GNU Multiple Precision Floating-Point Reliably 4.0.2, GNU NCURSES 6.1, GNU Readline 8.0, GNU Scientific Library 2.6, GNU Sed 4.7, GNU Tar 1.32, GNU Texinfo 4.7, GNU Wget 1.20.3, GNU Which 2.21, GPL Ghostscript 9.27, HDF5 library 1.10.5, ImageMagick 7.0.8-67, Imfit 1.6.1 \citep{imfit2015}, Libbsd 0.9.1, Libffi 3.2.1, Libgit2 0.28.2, Libjpeg v9b, Libpng 1.6.37, Libtiff 4.0.10, LibYAML 0.2.2, Lzip 1.20, Metastore (forked) 1.1.2-23-fa9170b, OpenBLAS 0.3.5, Open MPI 4.0.1, OpenSSL 1.1.1a, PatchELF 0.10, pkg-config 0.29.2, Python 3.7.4, SCons 3.0.5, SWarp 2.38.0 \citep{swarp}, Unzip 6.0, WCSLIB 6.4, XZ Utils 5.2.4, Zip 3.0 and Zlib 1.2.11.
Within Python, the following modules
were used: Astropy 3.2.1 \citep{astropy2013,astropy2018},  Cycler 0.10.0, Cython 0.29.6 \citep{cython2011},  h5py 2.9.0,  Kiwisolver 1.0.1, Matplotlib 3.1.1 \citep{matplotlib2007}, mpi4py 3.0.2 \citep{mpi4py2011}, Numpy 1.17.2 \citep{numpy2011}, pkgconfig 1.5.1,  PyParsing 2.3.1,  python-dateutil 2.8.0,  PyYAML 5.1,  Raulisfpy 0.1.2, Scipy 1.3.1 \citep{scipy2007,scipy2011},  Setuptools 40.8.0,  Setuptools-scm 3.2.0 and  Six 1.12.0.
The \LaTeX{} source of the paper was compiled
to make the PDF using the following packages:
biber 2.13, biblatex 3.13a, boondox 1.02d, caption 2019-09-01, courier 2016-06-24, csquotes 5.2e, datetime 2.60, ec 1.0, etoolbox 2.5h, fancyhdr 3.10, fmtcount 3.05, fontaxes 1.0d, footmisc 5.5b, fp 2.1d, kastrup 2016-06-24, logreq 1.0, newtx 1.607, pgf 3.1.4b, pgfplots 1.16, preprint 2011, setspace 6.7a, tex 3.14159265, texgyre 2.501, times 2016-06-24, titlesec 2.13, trimspaces 1.1, txfonts 2016-06-24, ulem 2016-06-24, xcolor 2.12 and xkeyval 2.7a. We are very grateful to all their
creators for freely providing this necessary
infrastructure. This research (and many 
others) would not be possible without them.

\bsp    
\label{lastpage}
\end{document}